\documentclass[showpacs, oneside, twocolumn, amsmath, amssymb,nofootinbib,superscriptaddress]{revtex4-1}

\usepackage{cases}
\usepackage{amsfonts}
\usepackage{dcolumn}
\usepackage{bm}
\usepackage{bbm}
\usepackage{graphicx}
\usepackage{xcolor}
\usepackage{array}
\usepackage{subfigure}
\usepackage{hyperref}
\usepackage{cancel}
\usepackage{empheq}

\newcommand{\be}{\begin{equation}}
\newcommand{\ee}{\end{equation}}
\newcommand{\ba}{\begin{eqnarray}}
\newcommand{\ea}{\end{eqnarray}}

\newcommand{\gsim}{\mathrel{\hbox{\rlap{\lower.55ex \hbox {$\sim$}}
                   \kern-.3em \raise.4ex \hbox{$>$}}}}
\newcommand{\lsim}{\mathrel{\hbox{\rlap{\lower.55ex \hbox {$\sim$}}
                   \kern-.3em \raise.4ex \hbox{$<$}}}}

\renewcommand{\v}[1]{\ensuremath{\mathbf{#1}}} 

\newcommand\R{\mathcal{R}}

\newcommand\A{\mathcal{A}}
\newcommand\cc{\mathcal{C}}
\newcommand\dd{\mathcal{D}}

\def\equationautorefname~#1\null{Eq.~#1\null}

\begin{document}

\centerline{DESY-19-120}
\title{A simple, exact, model of quasi-single field inflation} 

\author{Yvette Welling}%
 \email{yvette.welling@desy.de}
 \affiliation{%
 Deutsches Elektronen-Synchrotron DESY, Notkestra{\ss}e 85, 22607 Hamburg, Germany}

\date{\today}

\begin{abstract}
In this note we present a simple but exact model of quasi-single field inflation \cite{Chen:2009zp, Chen:2009we}, in which the couplings between perturbations are completely controlled, and for instance can be made constant with any desired value. This provides a way to numerically implement quasi-single field inflation and to test its predictions in various regimes. We confirm that the numerics agree to good accuracy with the analytical predictions for the isosceles bispectrum and the full shape function.
  
\end{abstract}

\maketitle

\section{Introduction}
Quasi-single field inflation \cite{Chen:2009zp, Chen:2009we} is a class of models where the inflaton interacts with other particles (isocurvatons) with mass $m\sim H$.  It predicts an interesting shape of primordial non-Gaussianities that is distinguishable from single-field inflation \cite{Chen:2009zp, Chen:2009we, Baumann:2011nk,  Assassi:2012zq, Noumi:2012vr, Chen:2014cwa,Arkani-Hamed:2015bza,Lee:2016vti, Arkani-Hamed:2018kmz}. The quasi-single field bispectrum has been constrained by CMB data \cite{Akrami:2018odb, Akrami:2019izv} and provides an important target for future observations (for a recent discussion see \cite{Meerburg:2016zdz}). Fields with mass of order Hubble may appear naturally in string embeddings of inflation, describing e.g. physics at the Kaluza-Klein scale \cite{Baumann:2014nda}. Moreover, masses of order Hubble could arise from radiative corrections to light fields \cite{Copeland:1994vg, Baumann:2011nk}.\\

Yet it is difficult to test the predictions of quasi-single field inflation with isocurvature mass $m\lesssim H$ numerically, because one needs to construct a potential that sustains a constantly curved trajectory in field space, while keeping $m$ under control. The centrifugal force displaces the inflaton from the minimum of the potential, leading to order Hubble corrections to the isocurvature mass. This problem was discussed in detail in \cite{Dias:2016rjq}. In this note we provide a simple model of quasi-single field inflation in which the couplings between perturbations are controlled, and for instance can be made constant with any desired value. \\

The main obstacle in these potential-based constructions is that, unlike single-field inflation, the potential $V(\varphi)$ does \textit{not} reflect the properties of the perturbations, because a curved inflationary trajectory deviates from the potential gradient flow in field space. Instead, we apply the recently developed orbital inflation \cite{Achucarro:2019pux, Achucarro:2019abc} to the quasi-single field regime. Orbital inflation is based on the Hamilton-Jacobi formalism developed in \cite{Muslimov:1990be, Salopek:1990jq,Lidsey:1991zp,Copeland:1993jj} and extended to multi-field in \cite{Skenderis:2006jq, Skenderis:2006rr,Byrnes:2009qy,Saffin:2012et,Garriga:2014fda, Garriga:2015tea, Achucarro:2018ngj} in which inflation attracts to the Hubble gradient flow
\begin{equation}
 \dot\phi^a \sim - \nabla^a H, 
\end{equation}
This means that the Hubble parameter $H(\phi^a)$ controls the behavior of both the background and perturbations.\\

The idea of orbital inflation is to align the inflationary trajectory - and thus the Hubble gradient - with an isometry of field space, 
i.e. an `angular' direction $\theta$ \cite{Achucarro:2019pux, Achucarro:2019abc}. This ensures that inflaton happens at constant `radius' $\rho = \rho_0$, like in the simplest models of quasi-single field inflation. Additionally, in a moment we will assume a product separable form of $H$ to meet the condition of constant isocurvature mass $m\sim H$. Therefore, the Hubble parameter - \textit{as opposed to} the potential - takes a particularly simple form to support quasi-single field inflation. Given the existence of an isometry we are free to write the kinetic term in the form
\begin{equation}
 -2K = f(\rho)(\partial \theta)^2 + (\partial \rho)^2\ .
\end{equation}
Slightly adapting\footnote{In the Hamilton-Jacobi formalism there is an additional contribution to the potential of $-2 H_{,\rho}^2$. However, on the isometric inflationary attractor this term is zero, and therefore we can preserve the same background solution if we remove this term. As pointed out in \cite{Achucarro:2018ngj} the Hamilton-Jacobi assumption bounds the isocurvature mass from above $\mu^2 \lesssim 3/2 H^2$, and to study quasi-single field inflation we would like to alleviate this constraint. Therefore, we modified the form of the potential.} the multi-field Hamilton-Jacobi formalism, the following potentials admit \textit{exact} inflationary attractors to the angular Hubble gradient flow 
\begin{equation}
 V=3H^2- \frac{2 H_{\theta}^2}{f(\rho)} \ ,
\end{equation}
where we denoted $H_\theta = \partial_\theta H$. 
Up to the constraint of the vanishing gradient in the radial direction $\partial_\rho H =0$, we are free to choose $H(\theta, \rho)$ and we can tune the couplings between perturbations to our liking. 
This is, to our knowledge, the first exact realization of quasi-single field inflation. \\

\section{The Model}\label{sec:modelofqsf}
Summarizing the discussion above, our simple model of quasi-single field inflation is given by
\begin{align}
 S_\phi = -\frac{1}{2} \int\ d^4 x\sqrt{-g}&\Bigg[R +  f(\rho)(\partial \theta)^2 + (\partial \rho)^2  \nonumber \\
 & + 6H^2(\theta,\rho)- \frac{4 H_{\theta}^2(\theta,\rho)}{f(\rho)}  \Bigg]\ ,
\end{align}  
with $H_{\rho}(\rho_0) =0 $ for some constant value $\rho_0$. Furthermore, $R$ is the Ricci scalar of spacetime and we use the convention $M_p = 1$. \\

Given some Hubble parameter $H(\theta, \rho)$ the relevant kinematical and geometrical inflationary quantities of this model are given by 
\begin{subequations}
\begin{empheq}{align}
& \frac{1}{\kappa} = \frac{f_\rho}{2 f}, \quad \dot\theta  = -2 \frac{H_\theta}{f}, \quad \epsilon  = \frac{2 H_\theta^2}{f H^2} , \\  
& \mathbb{R}=\frac{2}{\kappa^2}-\frac{f_{\rho\rho}}{f}, \quad
 m^2  = 6 H H_{\rho\rho}-\frac{4 H_\theta H_{\theta\rho\rho} }{f}\ , \label{eqn:entropymasssol} \\
 & \frac{V_{\rho\rho\rho}}{H^2} = \frac{6 H_{\rho\rho\rho}}{H} + \epsilon\left(\frac{12(2+\mathbb{R} \kappa)}{\kappa^3}+\frac{f_{\rho\rho\rho}}{f}\right) \nonumber \\
 & \qquad \quad  +\frac{4 H_\theta}{f H}\left(\frac{6 H_{\theta\rho\rho}}{H \kappa} - \frac{H_{\theta\rho\rho\rho}}{H}\right)\ .
 \label{eqn:Vrhorhorhosol}
\end{empheq}
\label{eqn:exactsolutionbg}
\end{subequations}
We use the shorthand notation $f=f(\rho)$, $H=H(\theta,\rho)$ and $H_\rho = \partial_\rho H$, etcetera.
Here $\kappa$ equals the field radius of curvature, $\epsilon \equiv  - \frac{\dot H}{H^2}$ is the first slow-roll parameter and $\mathbb{R}$ denotes the Ricci curvature of field space. Finally, $m$ is the effective mass of the radial isocurvature perturbations.  All variables are understood to be evaluated at $\rho=\rho_0$. \\

To reproduce the results of \cite{Chen:2009zp} we focus on a model of quasi-single field inflation with a constant isocurvature mass in Hubble. As can been seen from \autoref{eqn:entropymasssol}, the simplest way to realize a constant isocurvature mass and $V_{\rho\rho\rho}$ (up to a slow-roll correction) is to take a Hubble parameter of the product separable form 
\begin{equation}
H(\rho, \theta) = W(\theta)\left(1+ \frac{\lambda}{2} (\rho-\rho_0)^2 + \frac{\alpha}{6}(\rho-\rho_0)^3 + \ldots \right).\label{eqn:productseparablehubble}
\end{equation}

A product separable Hubble parameter yields the following subclass of potentials that admit quasi-single field inflation
\begin{align}
 &V(\theta, \rho) = 3 \left(W^2(\theta)-\frac{2 W_\theta^2(\theta)}{3f(\rho)}\right) \nonumber \\
 & \qquad \qquad \quad  \times \left(1+\frac{\lambda}{2}(\rho-\rho_0)^2+ \frac{\alpha}{6}(\rho-\rho_0)^3 + \ldots \right)^2\ .
 \label{eqn:reconstructedpotential}
\end{align}
Using \autoref{eqn:exactsolutionbg}, the isocurvature mass is given by $m^2/H^2 = 6\lambda + \mathcal{O}(\epsilon)$.
In addition, the higher order coupling that is important for non-gaussianity becomes $V_{\rho\rho\rho} = 6\alpha + \mathcal{O}(\epsilon)$. \\

Furthermore, we specialize to a flat field metric $f(\rho) =\rho^2$, such that $\kappa = \rho_0$ on the inflationary trajectory. Moreover, we take $W\sim \theta$, which corresponds to a quadratic type of potential along the angular direction with slow-roll parameter $\epsilon = 1/(2\Delta N +1)$.

\section{Numerical comparison}
We proceed by numerically checking the predictions of quasi-single field inflation. We use the PyTransport code developed by \cite{Mulryne:2016mzv, Ronayne:2017qzn} (see also \cite{Dias:2016rjq, Seery:2016lko}). In what follows the power spectrum and bispectrum are defined as
\begin{align}
&\langle \R(\v{k})\R(\v{p}) \rangle^\prime = P_\R(k) = \frac{2\pi^2}{k^3}\Delta^2_\R(k),\\
& \langle \R(\v{k}_1)\R(\v{k}_2)\R(\v{k}_3) \rangle^\prime =  B_\R(k_1,k_2,k_3)
\end{align}
where the prime denotes that the overal delta function $(2\pi)^3 \delta_D(\sum \v{k}_i)$ has been removed.
The bispectrum is usually expressed as a dimensionless shape function for which we use the following two conventions
\begin{equation}
 \mathcal{S}_\R(k_1,k_2,k_3) = \frac{k_1^2 k_2^2 k_3^2}{(2\pi^2)^2}\frac{B_\R(k_1,k_2,k_3)}{\Delta^2_\R(k_1) \Delta^2_\R(k_2) + \text{perm}}\ ,
\end{equation}
\begin{equation}
f_{NL}(k_1, k_2, k_3) \equiv \frac{5}{6}\frac{B_\R(k_1, k_2, k_3)}{P_\R(k_1) P_\R(k_2) + \text{perm}}\ . 
\end{equation}

An analytical expression for the power spectrum for orbital inflation - and hence of the model under consideration - was derived in \cite{Achucarro:2019abc}. The power spectrum is of the form
\begin{equation}
 \Delta^2_\R = \frac{H^2}{8\pi^2 \epsilon}(1+\mathcal{D}),
\end{equation}
where $\mathcal{D}$ captures the transfer of power from the isocurvature perturbations. 
Matching this result with the numerical function $\cc(\nu)$ defined in Equation 3.8 of \cite{Chen:2009zp}, we find 
\begin{equation}
 \cc(\nu) = \frac{\kappa^2}{16 \epsilon} \dd = \frac{1}{2}\left(\frac{1-\exp\left(\left(-\frac{3}{2}+\nu\right)\Delta N \right)}{\frac{3}{2}-\nu}\right)^2\ ,
\end{equation}
with $\nu \equiv \sqrt{9/4 - m^2/H^2}\approx \sqrt{9/4 - 6\lambda}$. 
This function agrees well with the one shown in Figure 6 of \cite{Chen:2009zp} if we take $\Delta N \sim 50-60$. On the other hand, for $\rho_0^2 \lesssim 10^2$ we don't find the same prediction for $n_s$ as in Eq. (3.11) of \cite{Chen:2009zp}, because the $\Delta N$-dependence of $C(\nu)$ becomes important.\\

\begin{figure}[h!]
\includegraphics[width=0.45\textwidth]{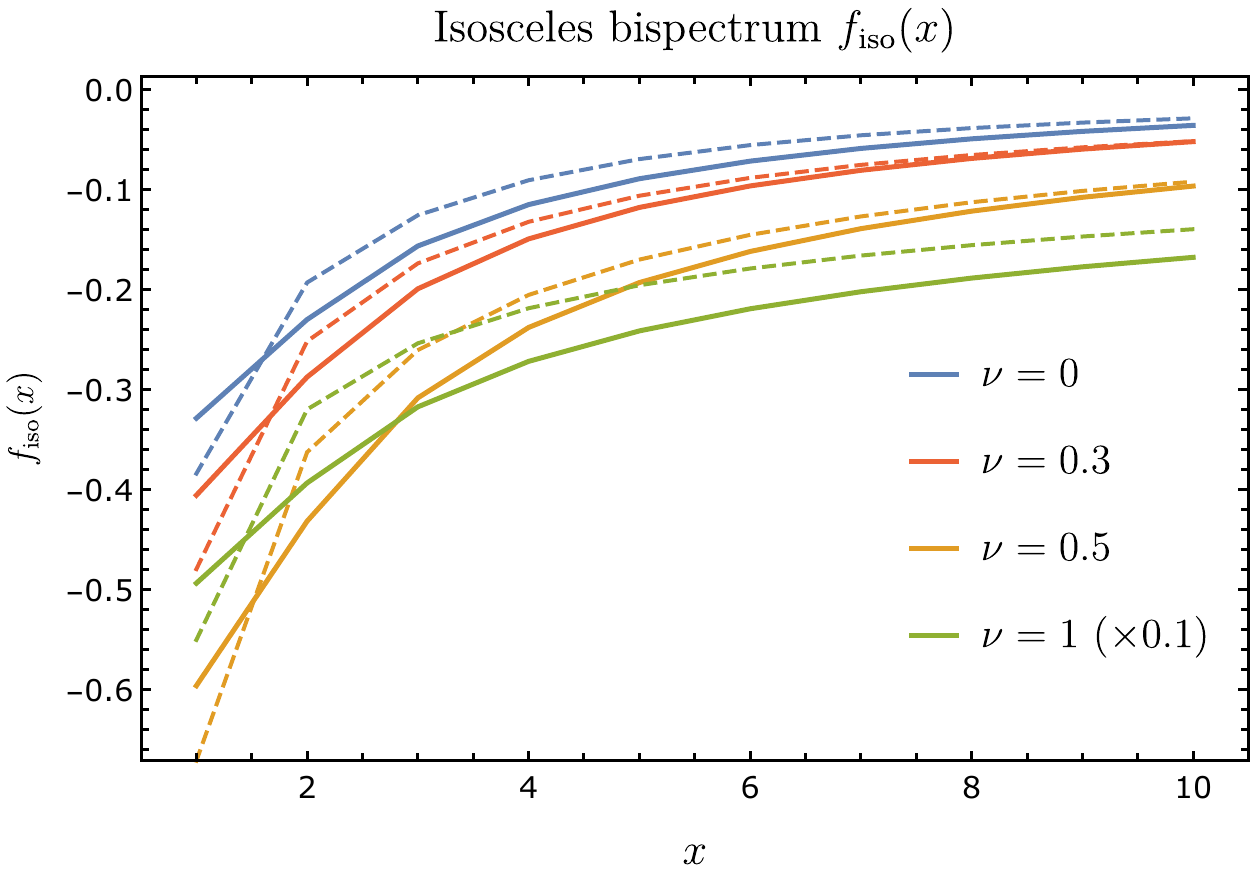}
\centering
\caption{This figure shows the analytical prediction $f_\text{ansatz}^\text{QSF}(x)$ in Equation \ref{eqn:crosssectionBSqsf} (dashed lines) and the full numerical solution (solid lines) of the isosceles bispectrum $f_\text{NL} (x)$ defined in Equation \ref{eqn:crosssectionBS} using the code \cite{Mulryne:2016mzv}. We take the values $\nu\in \{0.0, 0.3, 0.5, 1.0\}$. The amplitude of $f_\text{ansatz}^\text{QSF}(x)$ and the corresponding numerical solution for $\nu = 1$ have been rescaled by $0.1$ to fit inside the same plot range.}
\label{fig:crosssectionBS}
\end{figure}
Next, we compare the analytical predictions of the quasi-single field bispectrum \cite{Chen:2009zp} with the numerical solutions of our exact model. 
Perturbative control of the quasi-single field results requires 
$\left| \frac{V_{\rho\rho\rho}}{H}\right| \ll \left| \frac{V_{\rho\rho}}{H^2}\right|$,
which is amply satisfied if $\alpha \ll 1/H$. The analytical prediction of $f_\text{NL}$ in the equilateral configuration (Equation 4.16 of \cite{Chen:2009zp}) can then be written as 
\begin{equation}
 f^{\text{QSF}}_{\text{NL}} = -6\alpha H \A(\nu) (\Delta^2_\R)^{-1/2} \left(\frac{2\epsilon}{\rho_0^2}\right)^{3/2},
\end{equation}
with $\A(\nu)$ equal to $\alpha(\nu)$ defined in Equation 4.16 of \cite{Chen:2009zp}.
In the regime $\dd \ll 1$, that is assumed in \cite{Chen:2009zp}, we can approximate 
\begin{equation}
 f^{\text{QSF}}_{\text{NL}} \approx \A(\nu)  \frac{-48\alpha\pi \epsilon^2}{\rho_0^3}\ .
\end{equation}
Notice that $\alpha/\rho_0^3$ has to be quite large for the amplitude to be non slow-roll suppressed. At the same time we would like to stay in the regime $\dd \ll 1$.  
To compare with the results of \cite{Chen:2009zp} we therefore take $\alpha=1000$ and $\rho_0 =2$. \\

We start with comparing the shape function for isosceles configurations
\begin{equation}
f_\text{iso}(x) \equiv f_\text{NL}(k, xk, xk),
\label{eqn:crosssectionBS}
\end{equation}
where we suppress the $k$ dependence for convenience. 
Using the shape ansatz in Equation $6.2$ of \cite{Chen:2009zp}, the amplitude of the bispectrum for isosceles triangles is given by
\begin{equation}
  f^{\text{QSF}}_{\text{ansatz}}(x ) = \left(\frac{3}{2x+1}\right)^{7/2-3\nu}\frac{2x^2+1}{3x^{2\nu}}\times f^{\text{QSF}}_{\text{NL}}\ .
  \label{eqn:crosssectionBSqsf}
\end{equation}
In the numerical computation we vary $x$ between $1$ (equilateral configuration) and $x = 10$ (squeezed configuration). Moreover, we take $\nu\in \{0.0, 0.3, 0.5, 1.0\}$ and we take the long wavenumber to cross the horizon about 60 efolds before the end of inflation. The results are shown in Figure \ref{fig:crosssectionBS}. We see that the overall agreement is quite good. More precisely, we find that the predictions for the amplitude are accurate up to $30 \%$, $15 \%$, $20 \%$ and $25 \%$, respectively. \\

\begin{figure*}[h]
\includegraphics[width=0.95\textwidth]{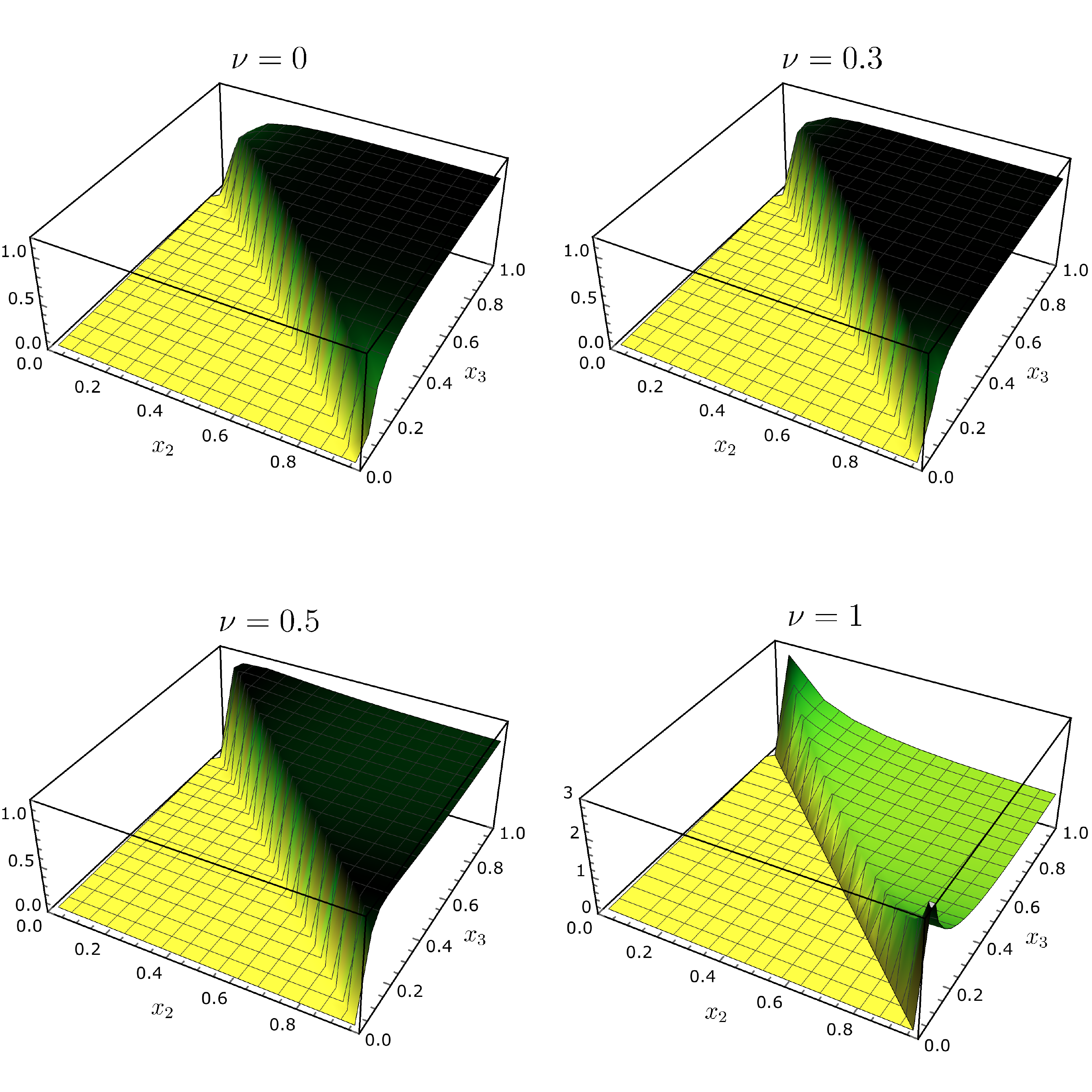}
\centering
\caption{The numerical solution for the shape function $\mathcal{S}(x_2, x_3)$ defined in \ref{eqn:shapefunctionS} using the code \cite{Mulryne:2016mzv}. We take the values $\nu\in \{0.0, 0.3, 0.5, 1.0\}$ and perform the necessarily rescalings, such that this Figure can be compared directly to Figure 7 of \cite{Chen:2009zp}.  }
\label{fig:shapefunctionsQSF}
\end{figure*}

\begin{figure*}[h]
\includegraphics[width=0.95\textwidth]{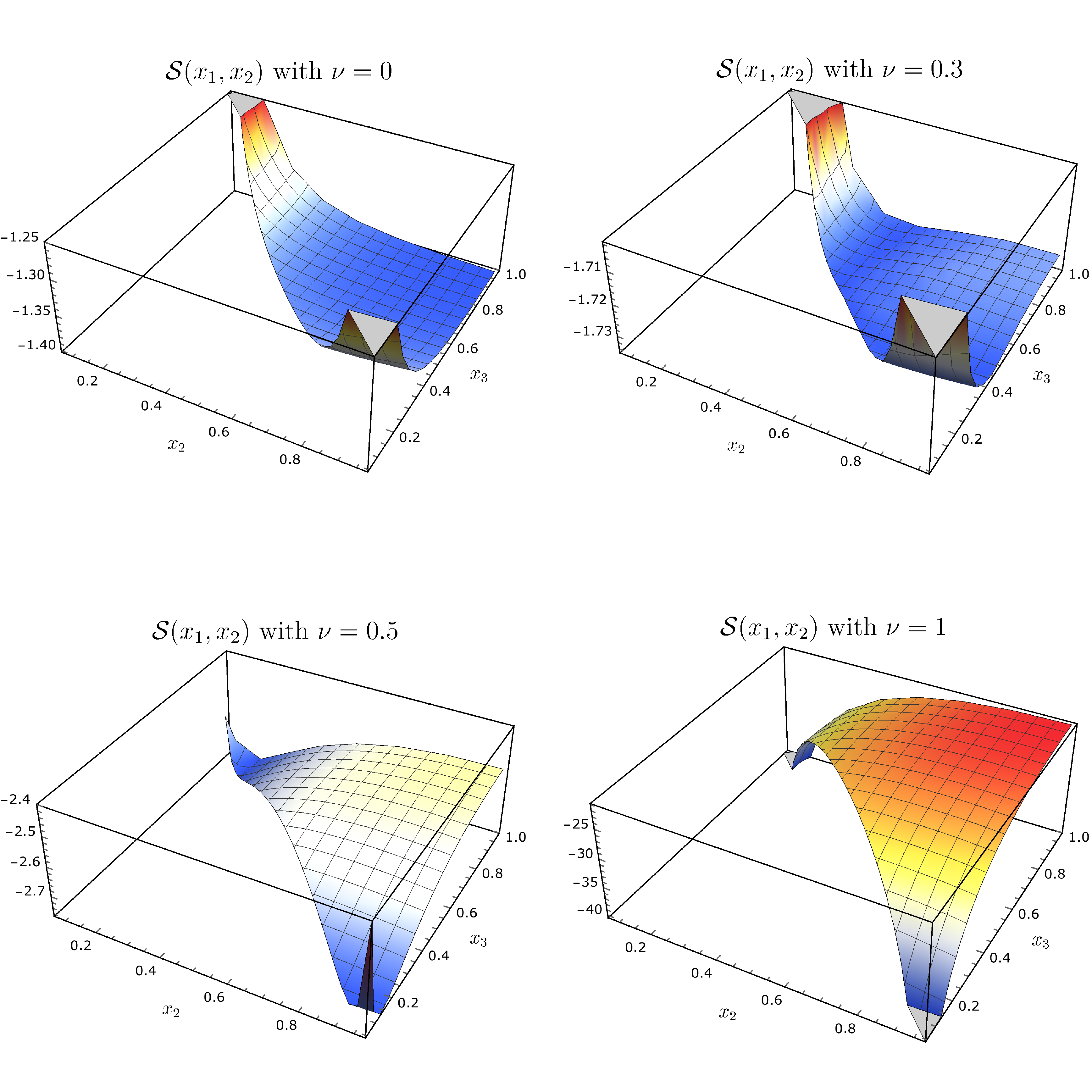}
\centering
\caption{The same as Figure \ref{fig:shapefunctionsQSF}, except that no rescalings of the shape function have been performed.}
\label{fig:shapefunctions}
\end{figure*}

Let us next compute the full shape function 
\begin{equation}
\mathcal{S}(x_1, x_2) \equiv \mathcal{S}_\R(k, x_2 k, x_3 k)\ ,
\label{eqn:shapefunctionS}
\end{equation}  
with the sum of the three wavevectors $k_t = k_1 + k_2 +k_3$ evaluated at 55 efolds before the end of inflation. The result is shown in Figure \ref{fig:shapefunctionsQSF}. To compare with Figure 7 of \cite{Chen:2009zp} we normalize the plot to one at the equilateral point $(x_2, x_3) = (1,1)$. Moreover we take a similar plot range. As far as we can judge the shapes are in agreement. For completeness we also show the unscaled shape functions in Figure \ref{fig:shapefunctions}, in which more detailed shape information can be seen.

\section{Conclusion}
In this note we provided a simple, exact, model of quasi-single field inflation. With this model we are able to test numerically the analytical predictions of \cite{Chen:2009zp}. We confirm that the numerics agree to good accuracy with the analytical predictions for the isosceles bispectrum and the full shape function.
 
\section{Acknowlegdments}
I wish to thank Ana Ach\'{u}carro for motivating me to write this note, and for stimulating discussions. Also, I would like to thank Gonzalo Palma and Dong-Gang Wang for discussions on related work.
YW is supported by the ERC Consolidator Grant STRINGFLATION under the HORIZON 2020 grant agreement no. 647995.

\bibliographystyle{hieeetr}
\bibliography{thesis}
\newpage

\end{document}